\documentclass [11pt,a4paper] {article}

\usepackage[T1]{fontenc}
\usepackage{lmodern}

\usepackage[cp1252]{inputenc}
\usepackage{amssymb}
\usepackage{amsmath}
\usepackage{amsfonts,amssymb}

\usepackage{graphicx}

\usepackage{bbm}
\usepackage{enumerate}

\usepackage[shortlabels]{enumitem}

\usepackage{amsthm}
\usepackage{cancel}

\usepackage{centernot}
\usepackage{mathrsfs}

\DeclareMathAlphabet{\mathpzc}{OT1}{pzc}{m}{it}

\def\SmallColSep{\setlength{\arraycolsep}{1pt}}

\newcommand*\rfrac[2]{{}^{#1}\!\!/\!_{#2}}

\newcommand{\notto}{\mathrel{{\ooalign{\hidewidth$\not\phantom{"}$\hidewidth\cr$\longleftrightarrow$}}}}

\begin{document}

\title{The paradox of classical reasoning}

\author{Arkady Bolotin\footnote{$Email: arkadyv@bgu.ac.il$\vspace{5pt}} \\ \emph{Ben-Gurion University of the Negev, Beersheba (Israel)}}

\maketitle

\begin{abstract}\noindent Intuitively, the more powerful a theory is, the greater the variety and quantity of ideas can be expressed through its formal language. Therefore, when comparing two theories concerning the same subject, it seems only reasonable to compare the expressive powers of their formal languages.\\

\noindent On condition that the quantum mechanical description is universal and so can be applied to macroscopic systems, quantum theory is required to be more powerful than classical mechanics. This implies that the formal language of Hilbert space theory must be more expressive than that of Zermelo–Fraenkel set theory (the language of classical formalism). However, as shown in the paper, such a requirement cannot be met. As a result, classical and quantum formalisms cannot be in a hierarchical relation, that is, include one another.\\

\noindent This fact puts in doubt the quantum-classical correspondence and undermines the reductionist approach to the physical world.\bigskip\bigskip

\bigskip\bigskip
\end{abstract}

\section{Introduction}  %{<-------------------------------------------------------------------------------------------------Section I}

\noindent Roughly speaking, the distinction between the macroscopic and the microscopic can be regarded as the difference between the classical and the quantum. Certainly, in many ways, classical mechanics can be considered a mainly macroscopic theory, whereas on the much smaller scale of atoms and molecules, classical mechanics may fail, and the interactions of particles are then described by quantum mechanics. Providing classical mechanics obeys bivalent truth-functional propositional logic (which is called \emph{classical logic}), one may expect that the domain of application of classical logic coincides with the macroscopic domain.\bigskip

\noindent Accordingly, assertions (i.e., declarative sentences or propositions) sensibly made about values of physical quantities possessed by \emph{a macroscopic system} can be handled using classical logic. For example, the assertion “The orbital angular momentum of a macroscopic body is \emph{up} along the axis of rotation of the body” can only have a bivalent truth value, true or false.\bigskip

\noindent By contrast, similar assertions made about values of physical quantities possessed by \emph{a quantum system} may violate classical logic. As an example, the declarative sentence “The spin of a spin-$\rfrac{1}{2}$ particle is \emph{up} along the $\overrightarrow{z\mkern8mu}$ axis” may have no bivalent truth value, i.e., be neither true nor false.\bigskip

\noindent Suppose that an experiment is conducted on a quantum system. Measurement outcomes of this experiment (such as a click in a detector or a position of a pointer) are specific values of physical quantities possessed by [a] macroscopic system[s] (i.e., [a] measurement device[s]) interacting with the observed quantum system. Hence, by the reason just given, sensible assertions about measurement outcomes of the experiment are expected to obey classical logic.\bigskip

\noindent Paradoxically, this intuition happens to be wrong: If a manipulating of sensible assertions about measurement outcomes of a quantum experiment is not allowed to produce a contradiction, then the belief that \emph{the logic of the macroscopic is classical logic} must be mistaken. Let us demonstrate this in the next section.\bigskip

\section{A thought experiment}  %{<-------------------------------------------------------------------------------------------------Section II}

\begin{figure}[ht!]
   \centering
   \includegraphics[scale=0.52]{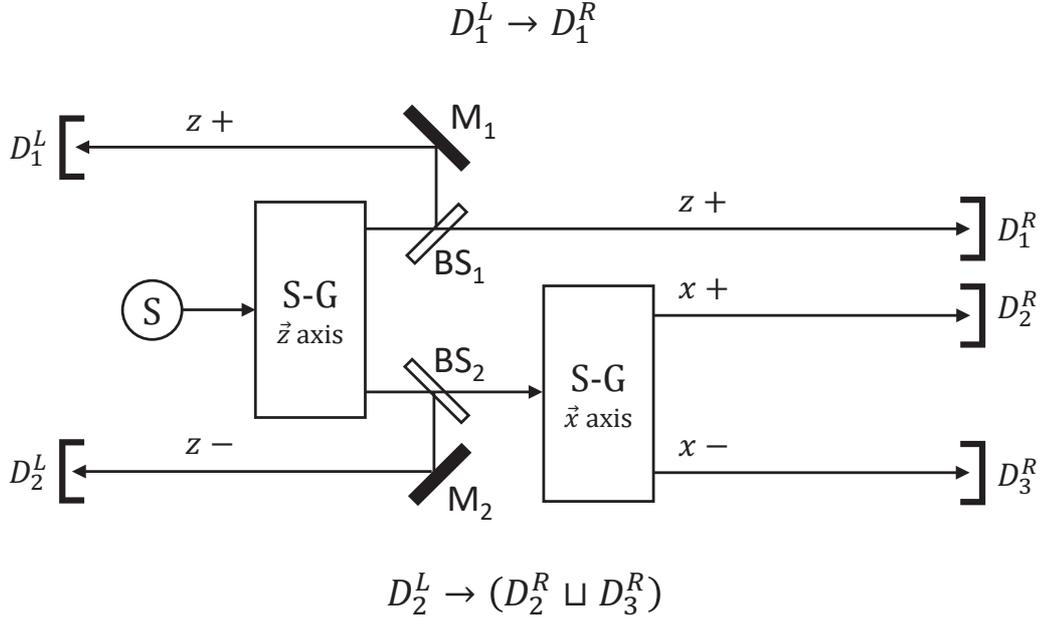}
   \caption{The experimental setup.\label{fig1}}
\end{figure}

\noindent Imagine the following experiment. The source S (see the Figure \ref{fig1}) sends electrically neutral spin-$\rfrac{1}{2}$ particles through the Stern-Gerlach magnet (see the rectangle containing the sign that reads “S-G $\overrightarrow{z\mkern8mu}$ axis”), which due to its orientation along the $\overrightarrow{z\mkern8mu}$ axis deflects the particles either up or down. At the end of the magnet, the beam of the particles whose spin component is pointing in the $+z$ direction (\emph{$z+$ beam} for short) is divided in two by the beam-splitter $\mathrm{BS_1}$. Equally, the particle beam with the spin component pointing in the $-z$ direction (\emph{$z-$ beam}) is divided in two by the beam-splitter $\mathrm{BS_2}$. Using mirrors $\mathrm{M_1}$ and $\mathrm{M_2}$, the reflected beams are redirected to a distant area on the left side of the experimental setup, where their arrivals are signaled (e.g., via clicking) by single-particle detectors (represented in the Figure \ref{fig1} as the thick left brackets).\bigskip

\noindent At the same time, the transmitted beams travel in the opposite direction such that one of them, namely, the $z-$ beam, passes through a further Stern-Gerlach magnet oriented along the $\overrightarrow{x\mkern8mu}$ axis (see the rectangle signed “S-G $\overrightarrow{x\mkern8mu}$ axis”). Finally, the transmitted $z+$ beam and the beams outputted from the second Stern-Gerlach magnet with the spin components aligned along the $\overrightarrow{x\mkern8mu}$ axis (\emph{$x+$ beam} and \emph{$x-$ beam}) reach an area situated at some distance away from the setup, where the additional detectors (depicted as the thick right brackets) can signal their arrivals.\bigskip 

\noindent To keep an account of what happens in the experiment \emph{interpretively neutral} (viz. to steer clear of the questions such as whether quantum mechanics is deterministic or stochastic, which elements of quantum mechanics can be considered real, and what the nature of measurement is), let us bring into play assertions sensibly made about measurement outcomes of the experiment.\bigskip

\noindent A description given by means of such assertions is bona fide \emph{instrumentalist}: It treats measurement as an unanalyzable act of verification (refutation) of the assertions. Most importantly, it describes only the effect of the observed quantum system on [a] macroscopic system[s], not the quantum system itself (for a general survey of instrumentalism see \cite{Timpson}).\bigskip

\noindent With that in mind, consider the following assertion: “The detector $n$ on the $s$ side of the experimental setup clicks”, where $s\in\{\text{left} ,\text{right}\}$ or $s\in\{L,R\}$ and $n\in\{1,2\}$ if $s=L$ or $n\in\{1,2,3\}$ if $s=R$. This assertion can be denoted by $D_n^s$.\bigskip

\noindent Because of the explanation given, verification of the declarative sentence $D_n^s$ (i.e., evidence that establishes or confirms the truth of $D_n^s$) refers to the factual click heard in the corresponding detector.\bigskip

\noindent The generalized uncertainty principle and the fact that in a situation involving incompatible measurement outcomes the occurrence of the one precludes the occurrence of the other[s] command against the simultaneous factual clicks on the same side of the experimental setup. Using symbols, this prohibition can be written in the form of a conjunction of two different sentences $D_n^s$, namely:\smallskip

\begin{equation} \label{QI} %{Eq.1}
   D_k^s
   \sqcap
   D_{m{\neq}k}^s
   \mkern-3mu
   \notto
   \mkern-4mu
   \top
   \;\;\;\;  ,
\end{equation}
\smallskip

\noindent where $\sqcap$ denotes the logical operator of conjunction, the left-right arrow with stroke stands for “is not equivalent to”, and $\top$ is an arbitrary tautology.\bigskip

\noindent Suppose that the sentence $D_1^L$ is verified. In that case, based on elementary facts about the properties of spin, one knows with absolute certainty that the sentence $D_1^R$ will be verified as well. Likewise, the verification of the sentence $D_2^L$ allows one to declare that either the sentence $D_2^R$ or the sentence $D_3^R$ is certain to be verified.\bigskip

\noindent This exposition can be expressed mathematically by means of the following conditional statements:\vspace{-8pt}

\begin{alignat}{3}
   \label{IMPL1} %{Eq.2}
   D_1^L
   \to
   &
   \mkern6mu
   D_1^R
   &
   \;\;\;\;  ,
\\[6pt]
   \label{IMPL2} %{Eq.3}
   D_2^L
   \to
   &
   \left(   
      D_2^R
      \sqcup
      D_3^R
   \right)   
   &
   \;\;\;\;  ,
\end{alignat}
\vspace{-8pt}

\noindent where $\to$ is used to denote material conditional, and $\sqcup$ denotes logical disjunction. Because of the property of the conditional, specifically, the conditional does not assume that its antecedent causes its consequent \cite{Klement}, the above statements do not specify a causal relationship between clicks in the detectors. That is, the statements make no claim that the click in either detector on the left side of the experimental setup acts at a distance on the detectors on the right side causing one of them click. Instead, these statements take to be the case that one knows which detector or detectors on the right side will click at the same time as one hears which detector on the left side clicks. It is worthy of notice that the propagation of knowledge such as this – whether or not it is faster than the speed of light – does not violate special relativity (the detailed explanation can be found, for example, in \cite{Samuel}).\bigskip

\noindent The conditionals (\ref{IMPL1}) and (\ref{IMPL2}) can be false only when $D_1^L$ and $D_2^L$ are true. On the other hand, as stated by the prohibition (\ref{QI}), the sentences $D_1^L$ and $D_2^L$ cannot be proved to be true at the same time. Hence, the conditionals (\ref{IMPL1}) and (\ref{IMPL2}) cannot be false together. Symbolically,\smallskip

\begin{equation} \label{EXP1} %{Eq.4}
   \neg
   \left(   
      D_1^L
      \to
      D_1^R
   \right)   
   \sqcap
   \neg
   \left(   
      D_2^L
      \to
      \left(   
         D_2^R
         \sqcup
         D_3^R
      \right)
   \right)
   \longleftrightarrow
   \bot
   \;\;\;\;  ,
\end{equation}
\smallskip

\noindent where $\neg$ means negation, $\longleftrightarrow$ stands for “is equivalent to”, and $\bot\longleftrightarrow\neg\top$ is an arbitrary contradiction. Providing De Morgan’s laws and double negation hold, the expression (\ref{EXP1}) can be rewritten as\smallskip

\begin{equation} \label{EXP2} %{Eq.5}
   \left(   
      D_1^L
      \to
      D_1^R
   \right)   
   \sqcup
    \left(   
      D_2^L
      \to
      \left(   
         D_2^R
         \sqcup
         D_3^R
      \right)
   \right)
   \longleftrightarrow
   \top
   \;\;\;\;  .
\end{equation}
\smallskip

\noindent Owing to Philo law\smallskip

\begin{equation} \label{PHILO} %{Eq.6}
   \left(   
      A
      \to
      B
   \right)   
   \longleftrightarrow
   \neg A
   \sqcup
   B
   \;\;\;\;  ,
\end{equation}
\smallskip

\noindent in which $A$ and $B$ are any sentences, from (\ref{EXP2}) it follows that\smallskip

\begin{equation}  %{Eq.7}
   \neg
   \left(   
      D_1^L
      \sqcap
      D_2^L
   \right)
   \sqcup
   \mkern-3mu  
   \left(   
      \mkern1mu
      D_1^R
      \sqcup
      \left(
         D_2^R
         \sqcup
         D_3^R
      \right)
   \right)   
   \longleftrightarrow
   \top
   \;\;\;\;  .
\end{equation}
\smallskip

\noindent Since $\neg\mkern2mu(D_1^L{\mkern3mu\sqcap\mkern3mu}D_2^L)\mkern-3mu\longleftrightarrow\mkern-3mu{\top}$, the propositional formula displayed above cannot assign any specific truth value to the disjunction $(\mkern1mu D_1^R \sqcup (D_2^R \sqcup D_3^R ))$, which can symbolically be expressed as follows:\smallskip

\begin{equation} \label{MAX} %{Eq.8}
   \left(   
      \mkern1mu
      D_1^R
      \sqcup
      \left(
         D_2^R
         \sqcup
         D_3^R
      \right)
   \right)   
   \notto
   \top
   \text{ or }
   \bot
   \;\;\;\;  .
\end{equation}
\smallskip

\noindent The foregoing expression states that before clicks are heard, assigning a truth value to each of the sentences $D_n^R$ can be done in a completely free manner, i.e., without being controlled or restricted by laws of the world and conditional solely upon one’s discretion. However, as long as an interpretation of the sentences concerning \emph{counterfactual clicks} (i.e., the ones that are not actually heard) stays free, the conjunctions $D_k^R{\mkern3mu\sqcap\mkern3mu}D_{m{\neq}k}^R$ may happen to be true.\bigskip

\noindent On the other hand, if a sentence is proved to be true (false) at some time, then this sentence must have been true (false) also at times when it was unproved. In another way, verification (refutation) merely reveals a pre-existing truth value of the sentence. On that account, simultaneous counterfactual clicks transpiring on the same side of the setup would contradict the prohibition (\ref{QI}).\bigskip

\noindent Hence, one gets a paradox: Even though the expression (\ref{MAX}) stems from the prohibition (\ref{QI}), the former contradicts the latter.\bigskip

\noindent For the most part, quantum paradoxes can be divided into two categories: those that challenge this or that axiom of quantum mechanics using classical logic and those that dispute this or that axiom of classical logic (non-modalized or otherwise) using quantum mechanics. Examples of paradoxes from the first category are Schrödinger's cat paradox and the EPR paradox calling into question the quantum mechanical principles of macroscopic realism and locality in that order \cite{Laloe, Scully}. The second category includes Hardy’s paradox \cite{Hardy} and the Frauchiger-Renner thought experiment \cite{Frauchiger} confronting the semantic principle of bivalence \cite{Bolotin} and the axioms of classical modal logic \cite{Nurgalieva, Boge}, respectively, in quantum settings.\bigskip

\noindent By contrast, the paradox presented here queries the ability of classical logic to refer to its own sentences when measurement outcomes of a quantum experiment are involved.\bigskip

\noindent Indeed, the line of argumentation used to arrive at this paradox is subject to the following assumptions:\vspace{-6pt}

\begin{description}[leftmargin=2.9cm]
\item[\rm{Assumption }\tt{A1}\rm{:}] \emph{Propositions are bivalent}. This means that sensible assertions about measurement outcomes of a quantum experiment are not allowed to take truth values over and above true and false.
\item[\rm{Assumption }\tt{A2}\rm{:}] \emph{Algebra of propositions is consistent}. This means that a manipulating of sensible assertions about measurement outcomes of a quantum experiment is not allowed to cause a contradiction.
\end{description}

\noindent Since the expression (\ref{MAX}) contradict the prohibition (\ref{QI}), one can conclude that the assumptions $\texttt{A1}$ and $\texttt{A2}$ cannot be true together. Symbolically, this can be presented using logical disjunction\smallskip

\begin{equation}  %{Eq.9}
   \left(
      \neg
      \texttt{A1}
      \sqcup
      \neg
      \texttt{A2}
   \right)
   \longleftrightarrow
   \top
   \;\;\;\;  ,
\end{equation}
\smallskip

\noindent or, equally, using material conditional\smallskip

\begin{equation}  %{Eq.10}
   \left(
      \texttt{A1}
      \to
      \neg
      \texttt{A2}
   \right)
   \longleftrightarrow
   \top
   \;\;\;\;  ,
\end{equation}
\vspace{-22pt}

\begin{equation}  %{Eq.11}
   \left(
      \texttt{A2}
      \to
      \neg
      \texttt{A1}
   \right)
   \longleftrightarrow
   \top
   \;\;\;\;  .
\end{equation}
\smallskip

\noindent In words, if classical logic is applicable to sensible assertions about measurement outcomes of a quantum experiment, then algebra of those assertions will be inconsistent, and vice versa. Thus, the paradox presented here can be called \emph{the paradox of classical reasoning}.\bigskip

\noindent Let us analyze how it can be resolved.\bigskip

\section{The Copenhagen interpretation}  %{<-------------------------------------------------------------------------------------------------Section III}

\noindent A way to solve the paradox would be to regard sensible assertions about measurement outcomes of a quantum experiment as being obedient to the rules of classical logic and seek additional postulates (apart from axioms of classical logic) that would make algebra of those propositions free from inconsistency.\bigskip

\noindent The Copenhagen interpretation propose a definite set of such postulates. Some of them (relevant to this discussion) are (see the full list in, e.g., \cite{Faye}):\vspace{-6pt}

\begin{description}[leftmargin=2.2cm]
\item[\rm{Postulate }\tt{A}\rm{.}] Quantum physics applies to individual objects, not only ensembles of objects.
\item[\rm{Postulate }\tt{B}\rm{.}] Their description is the result of experiments described in terms of classical physics.
\item[\rm{Postulate }\tt{C}\rm{.}] The ``frontier'' that separates the classical from the quantum can be chosen arbitrarily.
\item[\rm{Postulate }\tt{D}\rm{.}] No truth can be attributed to an object except according to the results of its measurement.
\end{description}

\noindent By pointing out that classical physics abides by the laws of classical logic, the postulate $\texttt{B}$ can be interpreted that every sensible assertion about factual clicks in the detectors should obey classical logic.\bigskip

\noindent Recall that before the clicks are heard in the experiment described above, the state of the spin-$\rfrac{1}{2}$ particles is the result of a superposition of two states, namely,\smallskip

\begin{equation}  %{Eq.12}
   |\Psi\rangle
   =
   \sum_{i=1}^2
   b_{i}
   |\Psi_{i}\rangle
   \;\;\;\;  ,
\end{equation}
\smallskip

\noindent where the summands are\smallskip

\begin{equation}  %{Eq.13}
   |\Psi_{i}\rangle
   =
   \left\{
      \begingroup\SmallColSep
      \begin{array}{l l}
         |\Psi_{z+}\rangle
         \otimes
         |\Psi_{z+}\rangle
         &
         \mkern-2mu
         ,
         \qquad
         i = 1
         \\[5pt]
         |\Psi_{z-}\rangle
         \otimes
         \left(
            c_1
            |\Psi_{x+}\rangle
            +
            c_2
            |\Psi_{x-}\rangle
         \right)
         &
         \mkern-2mu
         ,
        \qquad
         i = 2
         \\[5pt]
      \end{array}
      \endgroup   
   \right.   
   \;\;\;\;  ,
\end{equation}
\smallskip

\noindent and the complex coefficients $b_i$ and $c_i$ are such that $b_{i}^{\ast}b_{i}, c_{i}^{\ast}c_{i}\in[0,1]$.\bigskip

\noindent Now, in compliance with the concept of wavefunction collapse (required in the Copenhagen interpretation), assume that when either of the detectors on the left side clicks, the state $|\Psi\rangle$ “collapses” to the corresponding state $|\Psi_{i}\rangle$ that proves truth of the matching conditional. Explicitly, the state $|\Psi_{1}\rangle$ verifies the conditional $D_1^L{\mkern2mu\to\mkern2mu}D_1^R$, while the state $|\Psi_{2}\rangle$ verifies the conditional $D_2^L{\mkern2mu\to\mkern2mu}(D_2^R{\mkern2mu\sqcup\mkern2mu}D_3^R)$.\bigskip

\noindent Because of the property of collapse, be specific, the state $|\Psi\rangle$ collapses to just one of the states $|\Psi_{i}\rangle$, the said conditionals cannot be verified at the same time. This gives the formula additional to (\ref{EXP2}):\smallskip

\begin{equation}  %{Eq.14}
   \left(   
      D_1^L
      \to
      D_1^R
   \right)
   \sqcap
    \left(   
      D_2^L
      \to
      \left(
         D_2^R
         \mkern2mu
         \sqcup
         \mkern2mu
         D_3^R
      \right)
   \right)
   \longleftrightarrow
   \bot
   \;\;\;\;  .
\end{equation}
\smallskip

\noindent Applying the calculational device of classical logic, it produces the statement\smallskip

\begin{equation}  %{Eq.15}
   m
   \in
   \{2,3\}
   \mkern-3.3mu
   :
   \mkern10mu
   D_1^R
   \mkern4mu
   {\sqcap}
   \mkern4mu
   D_{m}^R
   \mkern4mu
   \longleftrightarrow
   \mkern6mu
   \bot
   \;\;\;\;  .
\end{equation}
\smallskip

\noindent In a similar fashion, when the state of the spin-$\rfrac{1}{2}$ particles subsequently collapses from the full $|\Psi_{2}\rangle$ to either $|\Psi_{z-}\rangle\otimes|\Psi_{x+}\rangle$ or $|\Psi_{z-}\rangle\otimes|\Psi_{x-}\rangle$, one gets the further formula\smallskip

\begin{equation}  %{Eq.16}
   \left(   
      D_2^L
      \to
      D_2^R
   \right)
   \sqcap
   \left(   
      D_2^L
      \to
      D_3^R
   \right)
   \longleftrightarrow
   \bot
   \;\;\;\;  
\end{equation}
\smallskip

\noindent that returns\smallskip

\begin{equation}  %{Eq.17}
   D_2^R
   \mkern4mu
   {\sqcap}
   \mkern4mu
   D_3^R
   \mkern4mu
   \longleftrightarrow
   \mkern6mu
   \bot
   \;\;\;\;  .
\end{equation}
\smallskip

\noindent Evidently, neither of the statements derived on account of wavefunction collapse violates the prohibition (\ref{QI}).\bigskip

\noindent As to the counterfactual clicks, declarative sentences about them must be meaningless in agreement with the postulate $\texttt{D}$. For example, the sentences $D_2^R$ and $D_3^R$ do not have the meaning up to the time of clicks in the corresponding detectors. In terms of formal semantics (i.e., the study of the meaning of sentences; see, for example, the reference book \cite{Fraassen}), this can be construed as that the sentences $D_2^R$ and $D_3^R$ have no truth value before collapse of the state $|\Psi_{2}\rangle$, even though afterward one of them becomes true and the other comes to be false.\bigskip

\noindent Be that as it may, let us inspect the sentences $D_2^R$ and $D_3^R$ a little further. If the meaning of the logical connectives $\neg$, $\sqcap$, and $\sqcup$ as well as the identity sign stay the same regardless of verification of their operands, then the complex sentences $\neg D_{2,3}^R$, $D_{2,3}^R \sqcap S$, $D_{2,3}^R \sqcup S$, and $D_{2,3}^R = S$ (where $S$ is an arbitrary sentence) must have no truth value before collapse, i.e., they must have something called \emph{truth-value gaps} \cite{Shaw}.\bigskip

\noindent Let us denote a truth-value gap by $u$ (short for “undefined'') and consider it as the third truth value that is not on a par with the others, i.e., $t$ (“true”) and $f$ (“false”). The special status of $u$ might be represented by the logical expressions (holding before collapse)\vspace{-8pt}

\begin{alignat}{3}
   \left(                        %{Eq.18}
      \neg D_{2,3}^R
   \right)
   &
   \longleftrightarrow
   u
   &
   \;\;\;\;  ,
\\[8pt]
   \left(                         %{Eq.19}
      D_{2,3}^R
      \mkern4mu
      {\sqcap}
      \mkern4mu
      S
   \right)
   &
   \longleftrightarrow
   u
   &
   \;\;\;\;  ,
\\[8pt]
   \left(                         %{Eq.20}
      D_{2,3}^R
      \mkern4mu
      {\sqcup}
      \mkern4mu
      S
   \right)
   &
   \longleftrightarrow
   u
   &
   \;\;\;\;  ,
\\[8pt]
   \left(                         %{Eq.21}
      D_{2,3}^R
      \mkern4mu
      {=}
      \mkern4mu
      S
   \right)
   &
   \longleftrightarrow
   u
   &
   \;\;\;\;  .
\end{alignat}
\vspace{-8pt}

\noindent Mathematically, these expressions are the same as they would be if they were gappy. Hence, a two-valued semantics with truth value gaps defines a logic that is identical to the one defined by a gapless three-valued semantics. This means that  the postulate $\texttt{D}$ negates the semantic principle of bivalence.\bigskip

\noindent Consequently, the Copenhagen interpretation is not consistent in obedience with classical logic. Namely, this interpretation declares the assumption $\texttt{A1}$ to be false before collapse but true afterward.\bigskip

\noindent The thing is that the Copenhagen interpretation does not elaborate on the nature of collapse. That is, according to the postulate $\texttt{C}$, collapse may be understood as a nonphysical process meaningful only in relation to a particular observer, e.g., one positioned on either side of the experimental setup (the case in point is Wigner’s friend thought experiment). As a result, the postulate $\texttt{B}$ (enabling the assumption $\texttt{A1}$) and the postulate $\texttt{D}$ (negating this assumption) may be applicable to the same sentence[s] simultaneously. In such a case, one would have a destructive inconsistency between $\texttt{B}$ and $\texttt{D}$ that can be presented in the form of the expression known as \emph{deductive explosion} \cite{Baskent}:\smallskip

\begin{equation}  %{Eq.22}
   \texttt{A1}
   ,
   \neg\texttt{A1}
   \vdash
   S
   \;\;\;\;  .
\end{equation}
\smallskip

\noindent Its wording is “If the assumption $\texttt{A1}$ and its negation both true, then it logically follows that any sentence $S$ is true”. For example, providing $S$ is $\bot$, the meaning of this expression is effectively $\top\vdash\bot$, which clearly cannot be valid.\bigskip

\noindent In a nutshell, from the inconsistency of the Copenhagen interpretation regarding classical logic, anything and everything can be concluded.\bigskip

\section{Quantum logic}  %{<-------------------------------------------------------------------------------------------------Section IV}

\noindent In contrast to the Copenhagen interpretation, one may firmly decide that classical logic is not the ``true'' logic. Accordingly, the belief that the logic of the macroscopic is classical logic must be considered unfounded.\bigskip

\noindent A version of propositional logic, which takes the principles of quantum theory into account and plays the role of the ``true'' logic, was proposed by Birkhoff and von Neumann in the 1936 paper \cite{Birkhoff} under the name of \emph{quantum logic}.\bigskip

\noindent As per quantum logic, logical disjunction is in general nondistributive over logical conjunction and vice versa \cite{Isham}, i.e.,\vspace{-8pt}

\begin{alignat}{5}
   A                              %{Eq.23}
   &
   \mkern4mu
   {\sqcap}_{q}
   \mkern4mu
   &
   \left(
      B
      \mkern4mu
      {\sqcup}_{q}
      \mkern4mu
      C
   \right)
   &
   \mkern4mu
   \notto
   \mkern6mu
   &
   \left(
      A
      \mkern4mu
      {\sqcap}_{q}
      \mkern4mu
      B
   \right)
   \mkern2mu
   {\sqcup}_{q}
   \mkern0.5mu
   \left(
      A
      \mkern4mu
      {\sqcap}_{q}
      \mkern4mu
      C
   \right)   
   \;\;\;\;  ,
\\[8pt]
   A                               %{Eq.24}
   &
   \mkern4mu
   {\sqcup}_{q}
   \mkern4mu
   &
   \left(
      B
      \mkern4mu
      {\sqcap}_{q}
      \mkern4mu
      C
   \right)
   &
   \mkern4mu
   \notto
   \mkern6mu
   &
   \left(
      A
      \mkern4mu
      {\sqcup}_{q}
      \mkern4mu
      B
   \right)
   \mkern2mu
   {\sqcap}_{q}
   \mkern0.5mu
   \left(
      A
      \mkern4mu
      {\sqcup}_{q}
      \mkern4mu
      C
   \right)   
   \;\;\;\;  ,
\end{alignat}
\vspace{-8pt}

\noindent where $A$, $B$ and $C$ are arbitrary non-logical symbols, whereas ${\sqcap}_{q}$ and ${\sqcup}_{q}$ are the quantum-logical connectives (counterparts of the classical connectives ${\sqcap}$ and ${\sqcup}$) that correspond to the infimum and supremum, respectively, of a lattice formed by closed subspaces of a Hilbert space ordered by set-inclusion (called \emph{the Hilbert lattice}).\bigskip

\noindent Consequently, in quantum logic, the material conditional defined by Philo law does not give rise to the unique implication connective. Surely, applying the fact that ${\neg}A{\sqcup}A$ is a tautology, the law (\ref{PHILO}) can be presented in two equivalent forms, namely,\smallskip

\begin{equation}  %{Eq.25}
   \left(   
      A
      \to
      B
   \right)   
   \longleftrightarrow
   \left(   
      \neg A
      \sqcup
      B
   \right)   
   \longleftrightarrow
   \left(   
      \neg A
      \sqcup
      \left(
         A
         \sqcap
         B
      \right)
   \right)   
   \;\;\;\;  .
\end{equation}
\smallskip

\noindent However, the above does not hold true in general if the quantum-logical connectives are used:\smallskip

\begin{equation}  %{Eq.26}
   \left(   
      {\neg}_{q}
      \mkern2mu
      A
      \mkern4mu
      {\sqcup}_{q}
      \mkern4mu
      B
   \right)   
   \notto
   \left(   
      {\neg}_{q}
      \mkern2mu
      A
      \mkern4mu
      {\sqcup}_{q}
      \mkern0.5mu
      \left(
         A
         \mkern4mu
         {\sqcap}_{q}
         \mkern4mu
         B
      \right)
   \right)   
   \;\;\;\;  ,
\end{equation}
\smallskip

\noindent where ${\neg}_{q}$, the quantum-logical counterpart of the classical negation $\neg$, corresponds to the operation of orthocomplement on the Hilbert lattice.\bigskip

\noindent So, unlike classical logic, quantum logic admits a set of different implications connectives. One of them that approximates the classical material conditional most fully is the conditional ${\to}_{1}$ (called \emph{Sasaki-hood}) \cite{Hardegree, Chiara}:\smallskip

\begin{equation} \label{HOOK} %{Eq.27}
   \left(   
      A
      {\to}_{1}
      B
   \right)   
   \longleftrightarrow
   \left(   
      {\neg}_{q}
      \mkern2mu
      A
      \mkern4mu
      {\sqcup}_{q}
      \mkern0.5mu
      \left(
         A
         \mkern4mu
         {\sqcap}_{q}
         \mkern4mu
         B
      \right)
   \right)   
   \;\;\;\;  .
\end{equation}
\smallskip

\noindent Using it, the quantum-logical counterpart of the expression (\ref{EXP2}) can be written down as\smallskip

\begin{equation} \label{QEXP2} %{Eq.28}
   \left(   
      D_1^L
      {\to}_{1}
      D_1^R
   \right)   
   \mkern2mu
   {\sqcup}_{q}
   \mkern0.5mu
    \left(   
      D_2^L
      \mkern2mu
      {\to}_{1}
      \mkern-2mu
      \left(   
         D_2^R
         \mkern4mu
         {\sqcup}_{q}
         \mkern4mu
         D_3^R
      \right)
   \right)
   \longleftrightarrow
   \top
   \;\;\;\;  .
\end{equation}
\smallskip

\noindent The statements derived therefrom are\vspace{-8pt}

\begin{alignat}{3}
   {\neg}_{q}                %{Eq.29}
   \mkern-2mu
   \left(
      D_1^L
      \mkern4mu
      {\sqcap}_{q}
      \mkern4mu
      D_2^L
   \right)
   &
   \longleftrightarrow
   \top
   &
   \;\;\;\;  ,
\\[8pt]
   \left(                         %{Eq.30}
      D_1^L
      \mkern4mu
      {\sqcap}_{q}
      \mkern4mu
      D_1^R
   \right)
   &
   \notto
   \top
   \text{ or }
   \bot
   &
   \;\;\;\;  ,
\\[8pt]
   D_2^L                         %{Eq.31}
   \mkern4mu
   {\sqcap}_{q}
   \mkern0.5mu
   \left(
      D_2^R
      \mkern4mu
      {\sqcup}_{q}
      \mkern4mu
      D_3^R
   \right)
   &
   \notto
   \top
   \text{ or }
   \bot
   &
   \;\;\;\;  .
\end{alignat}
\vspace{-8pt}

\noindent Neither contradicts the quantum-logical equivalent of the prohibition (\ref{QI}), i.e., $D_k^s\mkern4mu{\sqcap}_{\mkern-0.5mu q}\mkern4mu D_{m \neq k}^s\notto \top$. More than that, since $(D_1^L \mkern4mu{\sqcap}_{q}\mkern4mu D_1^R)\longleftrightarrow D_1^L$ and $(D_2^R \mkern4mu{\sqcup}_{q}\mkern4mu D_3^R) \longleftrightarrow \top$, the left-hand side of (\ref{QEXP2}) is always true\smallskip

\begin{equation}  %{Eq.32}
   \left(   
      {\neg}_{q}
      \mkern2mu
      D_1^L
      \mkern4mu
      {\sqcup}_{q}
      \mkern4mu
      D_1^L
   \right)   
   \mkern2mu
   {\sqcup}_{q}
   \mkern0.5mu
    \left(   
      {\neg}_{q}
      \mkern2mu
      D_2^L
      \mkern4mu
      {\sqcup}_{q}
      \mkern4mu
      D_2^L
   \right)
   \longleftrightarrow
   \top
   \;\;\;\;  ,
\end{equation}
\smallskip

\noindent and so (\ref{QEXP2}) is the equivalence $\top\longleftrightarrow\top$. Therefore, one can say that the paradox of classical reasoning does not take place within the confines of non-distributive logic (where both the assumption $\texttt{A1}$ and the assumption $\texttt{A2}$ are true).\bigskip

\noindent However, having said that, one must admit that the quantum connectives ${\sqcap}_{q}$ and ${\sqcup}_{q}$ can never be reduced to the classical operators of conjunction and disjunction. That is, under no condition is a non-distributive lattice such as the Hilbert lattice to become distributive, i.e., one that is isomorphic to a lattice of sets (closed under set union and intersection).\bigskip

\noindent On the other hand, even if non-distributive thinking had to be used to understand quantum phenomena, it would be still the case that distributive logic serves our daily life \cite{Putnam68}. This obvious remark leads to the hard question: How does distributive logic emerge from non-distributive one?\bigskip

\noindent The quick answer might be that we are ``duped'' in believing that there is a distributive logic so long as we only concern ourselves with macroscopic phenomena. But then another hard question arises: How can non-distributive logic explain the fact that formalism of quantum mechanics has been developed with distributive logic?\bigskip

\noindent Answering this question requires the completion of a difficult and presumably long program of ``recovering'' mathematics from quantum logic \cite{Dickson}. However, whether this program can be brought to fruition, or it is destined to fail, it is unknown.\bigskip

\noindent As appears, problems, which quantum logic gives rise to, outweigh the one it resolves.\bigskip

\section{Tarski's theory of truth}  %{<-------------------------------------------------------------------------------------------------Section V}

\noindent Instead of modifying the meaning of logical connectives that link up atomic sentences in the instrumentalist description – as does quantum logic – it is possible to modify the meaning of truth for those sentences. Let us examine the way in which this might be accomplished.\bigskip

\noindent As can be seen in the Section 2, the paradox of classical reasoning takes place only because some sentence[s] of the instrumentalist description, for example, the prohibition (\ref{QI}), is [are] allowed to predicate (i.e., affirm or assert) [a] truth value[s] of some other sentence[s] in the same description, e.g., the expression (\ref{MAX}).\bigskip

\noindent In accordance with Tarski's theory of truth \cite{Kirkham, Horsten}, to avoid self-contradiction (as the paradox of classical reasoning is), one must think about the hierarchy of languages wherein each language is able to predicate truth or falsehood of a sentence only in a language at a lower level. Then, if a sentence refers to the truth value of another sentence, the former will be semantically higher than the latter, preventing in this way self-contradiction.\bigskip

\noindent The sentence referred to is part of \emph{the object language}, while the referring sentence is a part of \emph{the metalanguage} with respect to the object language.\bigskip

\noindent In our case, the object language is the instrumentalist description of the physical world. By its name, this descriptionis is a formalized language (let us call it $\mathcal{L}_{\mkern3mu\mathrm{INS}}$). And what's more, $\mathcal{L}_{\mkern3mu\mathrm{INS}}$ lacks the universality of ordinary languages \cite{Priest}. To be precise, $\mathcal{L}_{\mkern3mu\mathrm{INS}}$ is \emph{not semantically closed}, which means that $\mathcal{L}_{\mkern3mu\mathrm{INS}}$ is not able to talk about its own semantics. Otherwise stated, $\mathcal{L}_{\mkern3mu\mathrm{INS}}$ cannot express the full range of semantic concepts that describe its own working. For this reason, to define the meaning of a given atomic sentence in $\mathcal{L}_{\mkern3mu\mathrm{INS}}$, that is, to explain when that sentence is true, one needs mathematical equations of a theory that deals with the physical world.\bigskip

\noindent So, it is natural to assume that the metalanguage pertaining to $\mathcal{L}_{\mkern3mu\mathrm{INS}}$ is a formal language of such a theory. In our case, this is the language of Hilbert space theory (let us call it $\mathcal{L}_{\mkern3mu\mathrm{HIL}}$) expressing the mathematical formulation of quantum mechanics (to be exact, its so-called Dirac’s abstract Hilbert space formalism) \cite{Edwards}. In particular, $\mathcal{L}_{\mkern3mu\mathrm{HIL}}$ expresses vectors of some Hilbert space along with their well-formed additions and scalar multiplications.\bigskip

\noindent In line with Tarski's truth condition \cite{Hodges}, the sentence $S\in\mathcal{L}_{\mkern3mu\mathrm{INS}}$ must be true if and only if its unique copy $\mathfrak{S}_S\in\mathcal{L}_{\mkern3mu\mathrm{HIL}}$ is true. In symbols, this can be written down as \emph{the T-sentence}:\smallskip

\begin{equation} \label{QTSEN} %{Eq.33}
   S
   \mkern-2mu\text{ is true in }\mkern-2mu
   \mathcal{L}_{\mkern3mu\mathrm{INS}}
   \iff
   \mathfrak{S}_S
   \mkern-2mu\text{ is true in }\mkern-2mu
   \mathcal{L}_{\mkern3mu\mathrm{HIL}}
   \;\;\;\;  ,
\end{equation}
\smallskip

\noindent where $\iff$ stands for ``if and only if''.\bigskip

\noindent A valid semantic theory must entail no more than one interpretive T-sentence for each sentence of the instrumentalist description \cite{Speaks}. Otherwise, an agent who knows all the theorems of the mathematical formalism will not yet understand the sentence $S$ because the agent will not know which of the T-sentences mentioning $S$ provides the meaning of $S$. This can be presented as the statement\smallskip

\begin{equation} \label{UNIQ} %{Eq.34}
   \forall
   S
   \mkern-2mu\in\mkern-2mu
   \mathcal{L}_{\mkern3mu\mathrm{INS}}
   \mkern-3.3mu
   :
   \mkern4mu
   \exists
   \mkern1.5mu
   !
   \mathfrak{S}_S
   \mkern-2mu\in\mkern-2mu
   \mathcal{L}_{\mkern3mu\mathrm{HIL}}
   \;\;\;\;  ,
\end{equation}
\smallskip

\noindent whose wordage is ``Given any sentence $S$ in $\mathcal{L}_{\mkern3mu\mathrm{INS}}$, that sentence must have exactly one copy $\mathfrak{S}_S$ in $\mathcal{L}_{\mkern3mu\mathrm{HIL}}$''.\bigskip

\noindent Thus, the ``technical'' problem is as follows: For a given sentence of the instrumentalist description $S$, find its unique counterpart $\mathfrak{S}_S$ in the language of Hilbert space theory.\bigskip

\noindent In the extreme cases of tautology and contradiction, this problem may have a simple solution, namely,\vspace{-8pt}

\begin{alignat}{3}
   &                          %{Eq.35}
   \top     
   \mkern-2mu\text{ is true in }\mkern-2mu
   \mathcal{L}_{\mkern3mu\mathrm{INS}}
   \iff
   |\Psi\rangle
   \mkern-2mu\in\mkern-2mu
   \mathcal{H}
   \mkern-2mu\text{ is true in }\mkern-2mu
   \mathcal{L}_{\mkern3mu\mathrm{HIL}}
   &
   \;\;\;\;  ,
\\[8pt]
   &                          %{Eq.36}
   \bot
   \mkern-2mu\text{ is true in }\mkern-2mu
   \mathcal{L}_{\mkern3mu\mathrm{INS}}
   \iff
   |\Psi\rangle
   \mkern-2mu\in\mkern-2mu
   \{0\}
   \mkern-2mu\text{ is true in }\mkern-2mu
   \mathcal{L}_{\mkern3mu\mathrm{HIL}}
   &
   \;\;\;\;  ,
\end{alignat}
\vspace{-8pt}

\noindent where $|\Psi\rangle$ is a unit vector that is associated with a definite (pure) state of a physical system, while $\mathcal{H}$ and $\{0\}$ are the identical and zero subspaces, respectively, of a Hilbert space of states. Equally, such cases can be interpreted using the following T-sentences:\vspace{-8pt}

\begin{alignat}{2}
   \top                   %{Eq.37}
   \mkern-2mu\text{ is true in }\mkern-2mu
   \mathcal{L}_{\mkern3mu\mathrm{INS}}
   \iff
   \hat{P}_{\top}
   |\Psi\rangle
   \mkern-3.5mu = \mkern-3.5mu
   |\Psi\rangle
   \mkern-2mu\text{ is true in }\mkern-2mu
   \mathcal{L}_{\mkern3mu\mathrm{HIL}}
   &
   \;\;\;\;  ,
\\[8pt]
   \bot                   %{Eq.38}
   \mkern-2mu\text{ is true in }\mkern-2mu
   \mathcal{L}_{\mkern3mu\mathrm{INS}}
   \iff
   \hat{P}_{\bot}
   |\Psi\rangle
   \mkern-3.5mu = \mkern-3.5mu
   |\Psi\rangle
   \mkern-2mu\text{ is true in }\mkern-2mu
   \mathcal{L}_{\mkern3mu\mathrm{HIL}}
   &
   \;\;\;\;  ,
\end{alignat}
\vspace{-8pt}

\noindent where $\hat{P}_{\top}=1$ and $\hat{P}_{\bot}=0$ are the projectors (i.e., self-adjoint idempotent operators on $\mathcal{H}$) which are in one-to-one correspondence with the subspaces $\mathcal{H}$ and $\{0\}$ such that $\mathrm{ran}(\hat{P}_{\top})=\mathcal{H}$ and $\mathrm{ran}(\hat{P}_{\bot})=\{0\}$ \cite{Mirsky}. Accordingly, the accurate formalization of truth for the sentence $S\in\mathcal{L}_{\mkern3mu\mathrm{INS}}$ is provided by the expression\smallskip

\begin{equation} \label{FORM} %{Eq.39}
   S
   \mkern-2mu\text{ is true in }\mkern-2mu
   \mathcal{L}_{\mkern3mu\mathrm{INS}}
   \iff
   \hat{P}_{S}
   |\Psi\rangle
   \mkern-3.5mu = \mkern-3.5mu
   |\Psi\rangle
   \mkern-2mu\text{ is true in }\mkern-2mu
   \mathcal{L}_{\mkern3mu\mathrm{HIL}}
   \;\;\;\;  ,
\end{equation}
\smallskip

\noindent where the formula $\hat{P}_{S}|\Psi\rangle\mkern-3.5mu = \mkern-3.5mu|\Psi\rangle$, evaluated to true for each given state $|\Psi\rangle$ of the system, is the copy of the sentence $S$ in the language of Hilbert space theory $\mathcal{L}_{\mkern3mu\mathrm{HIL}}$.\bigskip

\noindent As $\top\longleftrightarrow\neg\bot$, one gets:\vspace{-8pt}

\begin{alignat}{2}
   \neg                   %{Eq.40}
   \top
   \mkern-2mu\text{ is true in }\mkern-2mu
   \mathcal{L}_{\mkern3mu\mathrm{INS}}
   \iff
   0
   |\Psi\rangle
   \mkern-3.5mu = \mkern-3.5mu
   |\Psi\rangle
   \mkern-2mu\text{ is true in }\mkern-2mu
   \mathcal{L}_{\mkern3mu\mathrm{HIL}}
   &
   \;\;\;\;  ,
\\[8pt]
   \neg                   %{Eq.41}
   \bot
   \mkern-2mu\text{ is true in }\mkern-2mu
   \mathcal{L}_{\mkern3mu\mathrm{INS}}
   \iff
   1
   |\Psi\rangle
   \mkern-3.5mu = \mkern-3.5mu
   |\Psi\rangle
   \mkern-2mu\text{ is true in }\mkern-2mu
   \mathcal{L}_{\mkern3mu\mathrm{HIL}}
   &
   \;\;\;\;  ,
\end{alignat}
\vspace{-8pt}

\noindent which implies\smallskip

\begin{equation}  %{Eq.42}
   \neg
   S
   \mkern-2mu\text{ is true in }\mkern-2mu
   \mathcal{L}_{\mkern3mu\mathrm{INS}}
   \iff
   (1 - \hat{P}_{S})
   |\Psi\rangle
   \mkern-3.5mu = \mkern-3.5mu
   |\Psi\rangle
   \mkern-2mu\text{ is true in }\mkern-2mu
   \mathcal{L}_{\mkern3mu\mathrm{HIL}}
   \;\;\;\;  .
\end{equation}
\smallskip

\noindent Since $\top\sqcap\neg\top\longleftrightarrow\bot$, one may write\vspace{-8pt}

\begin{alignat}{2}
   &                                  %{Eq.43}
   \top\sqcap\neg\top
   \mkern-2mu\text{ is true in }\mkern-2mu
   \mathcal{L}_{\mkern3mu\mathrm{INS}}
   \iff
   \hat{P}_{\top}
   \hat{P}_{\bot}
   |\Psi\rangle
   \mkern-3.5mu = \mkern-3.5mu
   |\Psi\rangle
   \mkern-2mu\text{ is true in }\mkern-2mu
   \mathcal{L}_{\mkern3mu\mathrm{HIL}}
   &
   \;\;\;\;  ,
\\[8pt]
   &                                %{Eq.44}
   \top\sqcup\neg\top
   \mkern-2mu\text{ is true in }\mkern-2mu
   \mathcal{L}_{\mkern3mu\mathrm{INS}}
   \iff
   (\hat{P}_{\top} + \hat{P}_{\bot} - \hat{P}_{\top}\hat{P}_{\bot})
   |\Psi\rangle
   \mkern-3.5mu = \mkern-3.5mu
   |\Psi\rangle
   \mkern-2mu\text{ is true in }\mkern-2mu
   \mathcal{L}_{\mkern3mu\mathrm{HIL}}
   &
   \;\;\;\;  .
\end{alignat}
\vspace{-8pt}

\noindent On this account, if the projectors $\hat{P}_{S_1}$ and $\hat{P}_{S_2}$ commute, one can suggest that the following must hold:\smallskip

\begin{equation}  %{Eq.45}
   \hat{P}_{S_1}
   \hat{P}_{S_2}
   =
   \hat{P}_{S_2}
   \hat{P}_{S_1}
   \mkern-3.3mu
   :
   \mkern20mu
   \begingroup\SmallColSep
   \begin{array}{r l l}
      S_1\sqcap S_2
      \text{ is true in }
      \mathcal{L}_{\mkern3mu\mathrm{INS}}
      &
      \iff
      &
      \hat{P}_{ S_1\sqcap S_2}
      |\Psi\rangle
      \mkern-3.5mu = \mkern-3.5mu
      |\Psi\rangle
      \text{ is true in }
      \mathcal{L}_{\mkern3mu\mathrm{HIL}}
      \\[10pt]
      S_1\sqcup S_2
      \text{ is true in }
      \mathcal{L}_{\mkern3mu\mathrm{INS}}
      &
      \iff
      &
      \hat{P}_{ S_1\sqcup S_2}
      |\Psi\rangle
      \mkern-3.5mu = \mkern-3.5mu
      |\Psi\rangle
      \text{ is true in }
      \mathcal{L}_{\mkern3mu\mathrm{HIL}}      
   \end{array}
   \endgroup   
   \;\;\;\;  ,
\end{equation}
\smallskip

\noindent where $\hat{P}_{ S_1\sqcap S_2} \mkern-3.5mu = \mkern-3.5mu \hat{P}_{S_1}\mkern-2mu\wedge\mkern-2mu\hat{P}_{S_2} \mkern-3.5mu = \mkern-3.5mu \hat{P}_{S_1}\hat{P}_{S_2}$ and $\hat{P}_{ S_1\sqcup S_2} \mkern-3.5mu = \mkern-3.5mu \hat{P}_{S_1}\mkern-2mu\vee\mkern-2mu\hat{P}_{S_2} = \hat{P}_{S_1}+\hat{P}_{S_2} - \hat{P}_{S_1}\hat{P}_{S_2}$ are the lattice-theoretic meet and join, respectively, of the projectors $\hat{P}_{S_1}$ and $\hat{P}_{S_2}$.\bigskip

\noindent In case $\hat{P}_{S_1}$ and $\hat{P}_{S_2}$ do not commute, $\hat{P}_{ S_1\sqcap S_2}$ is neither $\hat{P}_{S_1}\hat{P}_{S_2}$ nor $\hat{P}_{S_2}\hat{P}_{S_1}$. By the same token, $\hat{P}_{ S_1\sqcup S_2}$ is not $\hat{P}_{S_1} + \hat{P}_{S_2} - \hat{P}_{S_1}\hat{P}_{S_2}$. Hence, in that case, $\hat{P}_{ S_1\sqcap S_2}$ and $\hat{P}_{ S_1\sqcup S_2}$ are elements of the sets containing all projectors $\hat{P}$ that satisfy the rules being defined prohibitively:\vspace{-8pt}

\begin{alignat}{4}
   \hat{P}_{S_1\sqcap S_2}  %{Eq.46}
   &
   \in
   \left\{
      \hat{P}
      \middle|
      \hat{P}
      \neq
      \hat{P}_{S_1}
      \hat{P}_{S_2}
      \text{ and }
      \hat{P}
      \neq
      \hat{P}_{S_2}
      \hat{P}_{S_1}
   \right\}
   &
   \;\;\;\;  ,
   \\[8pt]
   \hat{P}_{S_1\sqcup S_2}  %{Eq.47}
   &
   \in
   \left\{
      \hat{P}
      \middle|
      \hat{P}
      \neq
      \hat{P}_{S_1}
      +
      \hat{P}_{S_2}
      -
      \hat{P}_{S_2}
      \hat{P}_{S_1}
   \right\}
   &
   \;\;\;\;  .
\end{alignat}
\vspace{-8pt}

\noindent Then again, since neither $\hat{P}_{S_1}\hat{P}_{S_2}$ nor $\hat{P}_{S_2}\hat{P}_{S_1}$ is a projector therein, the above two sets can be described using the set-builder notation $\{\hat{P}|\hat{P}\text{ is not }(\text{not }\hat{P})\}$, which is equivalent to the set of all projectors $\{\hat{P}|\hat{P}=\hat{P}\} = \{\hat{P}\}$. This indicates that when $\hat{P}_{S_1}\hat{P}_{S_2}\neq\hat{P}_{S_2}\hat{P}_{S_1}$, the sentences $S_1\sqcap S_2$ and $S_1\sqcup S_2$ do not have unique copies $\hat{P}_{ S_1\sqcap S_2}|\Psi\rangle\mkern-3.5mu = \mkern-3.5mu |\Psi\rangle$ and $\hat{P}_{ S_1\sqcup S_2}|\Psi\rangle\mkern-3.5mu = \mkern-3.5mu |\Psi\rangle$ in $\mathcal{L}_{\mkern3mu\mathrm{HIL}}$; instead, their counterparts $\mathfrak{S}_{ S_1\sqcap S_2}$ and $\mathfrak{S}_{ S_1\sqcup S_2}$ in this metalanguage belong to the set holding formulas $\hat{P}|\Psi\rangle\mkern-3.5mu = \mkern-3.5mu |\Psi\rangle$ defined for each possible projector $\hat{P}$:\smallskip

\begin{equation} \label{NONC} %{Eq.48}
   \hat{P}_{S_1}
   \hat{P}_{S_2}
   \mkern-3.5mu\neq\mkern-3.5mu
   \hat{P}_{S_2}
   \hat{P}_{S_1}
   \mkern-3.3mu
   :
   \mkern2.5mu
   \begingroup\SmallColSep
   \begin{array}{r l l}
      S_1\sqcap S_2
      \mkern-2mu\text{ is true in }\mkern-2mu
      \mathcal{L}_{\mkern3mu\mathrm{INS}}
      &
      \mkern-2mu\iff\mkern-2mu
      &
      \mathfrak{S}_{ S_1\sqcap S_2}
      \mkern-2mu\in\mkern-2mu
      \left\{
         \hat{P}
         \middle|
         \hat{P}|\Psi\rangle\mkern-3.5mu = \mkern-3.5mu |\Psi\rangle
      \right\}
      \mkern-2mu\text{ is true in }\mkern-2mu
      \mathcal{L}_{\mkern3mu\mathrm{HIL}}      
      \\[8pt]
      S_1\sqcup S_2
      \mkern-2mu\text{ is true in }\mkern-2mu
      \mathcal{L}_{\mkern3mu\mathrm{INS}}
      &
      \mkern-2mu\iff\mkern-2mu
      &
      \mathfrak{S}_{ S_1\sqcup S_2}
      \mkern-2mu\in\mkern-2mu
      \left\{
         \hat{P}
         \middle|
         \hat{P}|\Psi\rangle\mkern-3.5mu = \mkern-3.5mu |\Psi\rangle
      \right\}
      \mkern-2mu\text{ is true in }\mkern-2mu
      \mathcal{L}_{\mkern3mu\mathrm{HIL}}
   \end{array}
   \endgroup
   \;\;\;\;  .
\end{equation}
\smallskip

\noindent Consequently, in that case, an agent who knows all the theorems of Hilbert space theory will not be able to understand the meaning of the sentences $S_1\sqcap S_2$ and $S_1\sqcup S_2$ of the instrumentalist description.\bigskip

\noindent Call to mind that a meaning of a sentence is its semantic value (i.e., truth value) \cite{Dever}. Hence, the expression (\ref{NONC}) indicates that the complex sentences $S_1\sqcap S_2$ and $S_1\sqcup S_2$ have truth-value gaps when $\hat{P}_{S_1}$ and $\hat{P}_{S_2}$ do not commute.\bigskip

\noindent Resultantly, the truth conditions of complex sentences cannot in general be reduced to the truth conditions of their constituents. So, as concerns the language of Hilbert space theory, complex sentences of the instrumentalist description are \emph{not truth-functional} in general.\bigskip

\noindent To be sure, let us assume that the opposite is true. Then, one can write\smallskip

\begin{equation}  %{Eq.49}
   S_1\sqcap S_2
   \text{ is true in }
   \mathcal{L}_{\mkern3mu\mathrm{INS}}
   \iff
   S_1
   \text{ is true in }
   \mathcal{L}_{\mkern3mu\mathrm{INS}}
   \text{ and }
   S_2
   \text{ is true in }
   \mathcal{L}_{\mkern3mu\mathrm{INS}}
   \;\;\;\;  .
\end{equation}
\smallskip

\noindent Combining this with (\ref{FORM}) one gets\smallskip

\begin{equation}  %{Eq.50}
   S_1\sqcap S_2
   \mkern-2mu\text{ is true in }\mkern-2mu
   \mathcal{L}_{\mkern3mu\mathrm{INS}}
   \mkern-6.5mu
   \iff
   \mkern-6mu
   \hat{P}_{S_1}
   |\Psi\rangle
   \mkern-3.5mu = \mkern-3.5mu
   |\Psi\rangle
   \mkern-2mu\text{ is true in }\mkern-2mu
   \mathcal{L}_{\mkern3mu\mathrm{HIL}}
   \text{ and }
   \hat{P}_{S_2}
   |\Psi\rangle
   \mkern-3.5mu = \mkern-3.5mu
   |\Psi\rangle
   \mkern-2mu\text{ is true in }\mkern-2mu
   \mathcal{L}_{\mkern3mu\mathrm{HIL}}
   \;\;\;\;  ,
\end{equation}
\smallskip

\noindent which implies the sentence\smallskip

\begin{equation}  %{Eq.51}
   S_1\sqcap S_2
   \text{ is true in }
   \mathcal{L}_{\mkern3mu\mathrm{INS}}
   \iff
   \hat{P}_{S_1}
   \hat{P}_{S_2}
   |\Psi\rangle
   \mkern-3.5mu = \mkern-3.5mu
   |\Psi\rangle
   \mkern-2mu\text{ is true in }\mkern-2mu
   \mathcal{L}_{\mkern3mu\mathrm{HIL}}
   \;\;\;\;   
\end{equation}
\smallskip

\noindent that is wrong in general: If $\hat{P}_{S_1}$ and $\hat{P}_{S_2}$ do not commute, the formula $\hat{P}_{S_1}\hat{P}_{S_2}|\Psi\rangle\mkern-3.5mu = \mkern-3.5mu|\Psi\rangle$ is not the copy of the sentence $S_1\sqcap S_2$ in $\mathcal{L}_{\mkern3mu\mathrm{HIL}}$.\bigskip

\noindent Modifying the meaning of truth does not break the distributive law of propositional logic. To see this, let us suppose that $\hat{P}_{S_1}\hat{P}_{S_1}\neq\hat{P}_{S_2}\hat{P}_{S_2}$ and then consider the following T-sentences:\vspace{-8pt}

\begin{alignat}{3}
   S_1             %{Eq.52}
   \sqcup
   \left(
      S_2
      \sqcap
      \neg S_2
   \right)
   \mkern-2mu\text{ is true in }\mkern-2mu
   \mathcal{L}_{\mkern3mu\mathrm{INS}}
   &
   \iff
   \hat{P}_{S_1}
   |\Psi\rangle
   \mkern-3.5mu = \mkern-3.5mu
   |\Psi\rangle
   \mkern-2mu\text{ is true in }\mkern-2mu
   \mathcal{L}_{\mkern3mu\mathrm{HIL}}
   &
   \;\;\;\;   ,
   \\[8pt]
   \left(             %{Eq.53}
      S_1
      \sqcup
      S_2
   \right)
   \sqcap
   \left(
      S_2
      \sqcup
      \neg S_2
   \right)
   \mkern-2mu\text{ is true in }\mkern-2mu
   \mathcal{L}_{\mkern3mu\mathrm{INS}}
   &
   \iff
   \mathfrak{S}_{ S_1\sqcup S_2}
   \mkern-2mu\in\mkern-2mu
   \left\{
      \hat{P}
      \middle|
      \hat{P}|\Psi\rangle\mkern-3.5mu = \mkern-3.5mu |\Psi\rangle
   \right\}
   \mkern-2mu\text{ is true in }\mkern-2mu
   \mathcal{L}_{\mkern3mu\mathrm{HIL}}
   &
   \;\;\;\;   .
\end{alignat}
\vspace{-8pt}

\noindent Using the gapless three-valued semantics mentioned in the Section 3, one finds\smallskip

\begin{equation}  %{Eq.54}
   \bigg(
      \Big(
         S_1
         \sqcup
         \left(
            S_2
            \sqcap
            \neg S_2
          \right)
      \Big)
      =
      \Big(
         \left(
            S_1
            \sqcup
            S_2
         \right)
         \sqcap
         \left(
            S_2
            \sqcup
            \neg S_2
         \right)
      \Big)
   \bigg)
   \longleftrightarrow
   u
   \;\;\;\;   .
\end{equation}
\smallskip

\noindent As the overhead expression indicates, there is no point in stating that the meanings of the propositions positioned on the opposite sides of the identity sign are different. In other words, the overhead expression does not necessitate the failure of distributivity.\bigskip

\noindent For this reason, the formalization of a true conditional $S_1 \to S_2$ can be given by the following T-sentence:\smallskip

\begin{equation}  %{Eq.55}
   \left(
      S_1
      \to
      S_2
   \right)
   \mkern-2mu\text{ is true in }\mkern-2mu
   \mathcal{L}_{\mkern3mu\mathrm{INS}}
   \iff
   \hat{P}_{\neg S_1 \sqcup S_2}
   |\Psi\rangle
   \mkern-3.5mu = \mkern-3.5mu
   |\Psi\rangle
   \mkern-2mu\text{ is true in }\mkern-2mu
   \mathcal{L}_{\mkern3mu\mathrm{HIL}}
   \;\;\;\;   .
\end{equation}
\smallskip

\noindent Let us review how the modification of the meaning of truth can resolve the paradox of classical reasoning.\bigskip

\noindent According to this modification, saying that the sentence $D_{n}^{s}$ ``is true'' must mean the following:\smallskip

\begin{equation}  %{Eq.56}
   D_{n}^{s}
   \text{ is true in }
   \mathcal{L}_{\mkern3mu\mathrm{INS}}
   \iff
   \hat{P}_{D_{n}^{s}}
   |\Psi\rangle
   \mkern-3.5mu = \mkern-3.5mu
   |\Psi\rangle
   \mkern-2mu\text{ is true in }\mkern-2mu
   \mathcal{L}_{\mkern3mu\mathrm{HIL}}
   \;\;\;\;  ,
\end{equation}
\smallskip

\noindent where the projectors $\hat{P}_{D_{n}^{s}}$ can be expressed through the eigenvectors of the Pauli matrices \cite{Piron}:\vspace{-8pt}

\begin{alignat}{3}
   \hat{P}_{D_{1}^{L,R}}  %{Eq.57}
   &
   =
   &
   \left|
      \frac{1}{2}
      ,
   \right.
   \left.
      +\frac{1}{2}
   \right\rangle_{\mkern-8mu z}
   \mkern-2mu
   \left\langle
      \frac{1}{2}
      ,
   \right.
   \left.
      +\frac{1}{2}
   \right|_{\mkern-2mu z}
   \;\;\;\;  ,
\\[8pt]
   \hat{P}_{D_{2}^{L}}  %{Eq.58}
   &
   =
   &
   \left|
      \frac{1}{2}
      ,
   \right.
   \left.
      -\frac{1}{2}
   \right\rangle_{\mkern-8mu z}
   \mkern-2mu
   \left\langle
      \frac{1}{2}
      ,
   \right.
   \left.
      -\frac{1}{2}
   \right|_{\mkern-2mu z}
   \;\;\;\;  ,
\\[8pt]
   \hat{P}_{D_{2}^{R}}  %{Eq.59}
   &
   =
   &
   \left|
      \frac{1}{2}
      ,
   \right.
   \left.
      +\frac{1}{2}
   \right\rangle_{\mkern-8mu x}
   \mkern-2mu
   \left\langle
      \frac{1}{2}
      ,
   \right.
   \left.
      +\frac{1}{2}
   \right|_{\mkern-2mu x}
   \;\;\;\;  ,
\\[8pt]
   \hat{P}_{D_{3}^{R}}  %{Eq.60}
   &
   =
   &
   \left|
      \frac{1}{2}
      ,
   \right.
   \left.
      -\frac{1}{2}
   \right\rangle_{\mkern-8mu x}
   \mkern-2mu
   \left\langle
      \frac{1}{2}
      ,
   \right.
   \left.
      -\frac{1}{2}
   \right|_{\mkern-2mu x}
   \;\;\;\;  .
\end{alignat}
\vspace{-8pt}

\noindent As is apparent,\vspace{-8pt}

\begin{alignat}{6}
   &
   \hat{P}_{D_{1}^{L}}           %{Eq.61}
   \hat{P}_{D_{2}^{L}}
   &
   \mkern4mu
   =
   &
   \mkern4mu
   \hat{P}_{D_{2}^{L}}
   \hat{P}_{D_{1}^{L}}
   &
   =
   &
   \mkern8mu
   0
   &
   \;\;\;\;  .
\\[8pt]
   m                                      %{Eq.62}
   \in
   \{2,3\}
   \mkern-3.3mu
   :
   \mkern10mu
   &
   \hat{P}_{D_{1}^{R}}
   \hat{P}_{D_{m}^{R}}
   &
   \mkern4mu
   \neq
   &
   \mkern4mu
   \hat{P}_{D_{m}^{R}}
   \mkern-4mu
   \hat{P}_{D_{1}^{R}}
   &
   \text{ }
   &
   \mkern8mu
   \text{ }
   &
   \;\;\;\;  ,
\\[8pt]
   &
   \hat{P}_{D_{2}^{R}}           %{Eq.63}
   \hat{P}_{D_{3}^{R}}
   &
   \mkern4mu
   =
   &
   \mkern4mu
   \hat{P}_{D_{3}^{R}}
   \hat{P}_{D_{2}^{R}}
   &
   =
   &
   \mkern8mu
   0
   &
   \;\;\;\;  .
\end{alignat}
\vspace{-8pt}

\noindent Consequently, the T-sentence corresponding to the expression (\ref{MAX}) is\smallskip

\begin{equation} \label{T1} %{Eq.64}
   \left(   
      \mkern1mu
      D_1^R
      \sqcup
      \left(
         D_2^R
         \sqcup
         D_3^R
      \right)
   \right)   
   \mkern-2mu\text{ is true in }\mkern-2mu
   \mathcal{L}_{\mkern3mu\mathrm{INS}}
   \iff
   \hat{P}_{\top}|\Psi\rangle
   \mkern-3.5mu
   =
   \mkern-3.5mu
   |\Psi\rangle
   \mkern-2mu\text{ is true in }\mkern-2mu
   \mathcal{L}_{\mkern3mu\mathrm{HIL}}
   \;\;\;\;  .
\end{equation}
\smallskip

\noindent Meanwhile, the T-sentences associated with the prohibition of simultaneous counterfactual clicks (that would have happened) on the right side of the setup are:\vspace{-8pt}

\begin{alignat}{5}
   m                             \label{T2} %{Eq.65}
   \mkern-2mu\in\mkern-2mu
   \{2,3\}
   \mkern-3.3mu
   :
   \mkern5mu
   &
   D_1^R
   \sqcap
   D_m^R
   \mkern-0.5mu\text{ is true in }\mkern-2mu
   \mathcal{L}_{\mkern3mu\mathrm{INS}}
   &
   \iff
   &
   \mathfrak{S}_{D_1^R \sqcap D_m^R}
   \mkern-2mu\in\mkern-2mu
   \left\{
      \hat{P}
      \middle|
      \hat{P}|\Psi\rangle\mkern-3.5mu = \mkern-3.5mu |\Psi\rangle
   \right\}
   \mkern-2mu\text{ is true in }\mkern-2mu
   \mathcal{L}_{\mkern3mu\mathrm{HIL}}
   &
   \;\;\;\;  ,
\\[8pt]
   &                             \label{T3} %{Eq.66}
   D_2^R
   \sqcap
   D_3^R
   \mkern-0.5mu\text{ is true in }\mkern-2mu
   \mathcal{L}_{\mkern3mu\mathrm{INS}}
   &
   \iff
   &
   \hat{P}_{\bot}|\Psi\rangle
   \mkern-3.5mu
   =
   \mkern-3.5mu
   |\Psi\rangle
   \mkern-2mu\text{ is true in }\mkern-2mu
   \mathcal{L}_{\mkern3mu\mathrm{HIL}}
   &
   \;\;\;\;  .
\end{alignat}
\vspace{-8pt}

\noindent Based on the above and using once again the gapless three-valued semantics, one can compose the following expressions:\vspace{-8pt}

\begin{alignat}{5}
   \bigg(                        %{Eq.67}
      \mkern-4mu
      \left(   
         \mkern1mu
         D_1^R
         \sqcup
         \left(
            D_2^R
            \sqcup
            D_3^R
         \right)
      \right)   
      =
      &
      \left(   
         \mkern1mu
         D_1^R
         \sqcup
         \left(
            D_2^R
            \sqcap
            D_3^R
         \right)
      \right)   
      \mkern-4mu
   \bigg)
   \longleftrightarrow
   f
   &
   \;\;\;\;  ,
\\[4pt]
   \bigg(                        %{Eq.68}
      \mkern-4mu
      \left(   
         \mkern1mu
         D_1^R
         \sqcup
         \left(
            D_2^R
            \sqcup
            D_3^R
         \right)
      \right)   
      =
      &
      \left(   
         \left(
            D_1^R
            \sqcap
            D_{m \neq 1}^R
         \right)
         \sqcap
         D_{k \neq m}^R
      \right)
      \mkern-4mu
   \bigg)
   \longleftrightarrow
   u
   &
   \;\;\;\;  .
\end{alignat}
\vspace{-8pt}

\noindent According to them, the propositions enclosed in the outermost parentheses are either false or undefined but never true. This means that the truth of the disjunction $(\mkern1mu D_1^R \sqcup (D_2^R \sqcup D_3^R ))$ never contradicts the prohibition of simultaneous counterfactual clicks on the right side of the setup. Consequently, when the meaning of truth for the sentences $D_n^s$ is modified, their manipulation within the rules of classical logic does not deduce a contradiction. Ergo, the assumption $\texttt{A2}$ holds true.\bigskip

\noindent On the other hand, because not all sentences of $\mathcal{L}_{\mkern3mu\mathrm{INS}}$ can be expressed uniquely in $\mathcal{L}_{\mkern3mu\mathrm{HIL}}$, the instrumentalist description of a quantum experiment can only be partially interpreted. Formally, this may be presented as the antithesis of the statement (\ref{UNIQ})\smallskip

\begin{equation} %{Eq.69}
   \exists
   \mkern2mu
   S
   \mkern-2mu\in\mkern-2mu
   \mathcal{L}_{\mkern3mu\mathrm{INS}}
   \mkern-3.3mu
   :
   \mkern1.5mu
   \exists
   \mkern2mu
   \mathfrak{S}_S
   ,
   \mathfrak{T}_S
   \mkern-2mu\in\mkern-2mu
   \mathcal{L}_{\mkern3mu\mathrm{HIL}}
   \;\;\;\;   
\end{equation}
\smallskip

\noindent that declares, ``For at least one sentence $S$ of $\mathcal{L}_{\mkern3mu\mathrm{INS}}$, there are at least two different copies $\mathfrak{S}_S$ and $\mathfrak{T}_S$ in $\mathcal{L}_{\mkern3mu\mathrm{HIL}}$''.\bigskip

\noindent It, then, should be concluded that the instrumentalist description, as expressed by the language of Hilbert space theory, fails to obey classical logic: Some of its sentences do not have the meanings that make them either true or false. This proves the assumption $\texttt{A1}$ to be mistaken.\bigskip

\section{The quantum-classical correspondence}  %{<-------------------------------------------------------------------------------------------------Section VI}

\noindent Let us consider assertions sensibly made about values of physical quantities possessed by [a] macroscopic system[s], as for example “The orbital angular momentum of a macroscopic body is \emph{up} along the axis of rotation of the body”. Providing such assertions can be regarded as \emph{instruments} used to explain and predict macroscopic phenomena, the description through them can be called the instrumentalist description at the macroscopic scale.\bigskip

\noindent In this case, a metalanguage with respect to the instrumentalist description $\mathcal{L}_{\mkern3mu\mathrm{INS}}$ is expected to be a formal language of classical mechanics $\mathcal{L}_{\mkern3mu\mathrm{CM}}$. Based on the concept of phase space of a classical system (in which all possible states of the system are represented such that each corresponds to exactly one point $q$ in the phase space $\Gamma$), the said language can be identified as a subset of $\mathcal{L}_{\mkern3mu\mathrm{ZF}}$, a formal language of Zermelo-Frankel set theory (ZF) \cite{Arnold}. Symbolically,\smallskip

\begin{equation} %{Eq.70}
   \mathcal{L}_{\mkern3mu\mathrm{CM}}
   \subseteq
   \mathcal{L}_{\mkern3mu\mathrm{ZF}}
   \;\;\;\;    ,
\end{equation}
\smallskip

\noindent which means that all well-formed formulas (WFFs for short), which can be written in $\mathcal{L}_{\mkern3mu\mathrm{CM}}$, can also be written in $\mathcal{L}_{\mkern3mu\mathrm{ZF}}$ (with or without transformations that account for the difference in the grammar of the languages). Hence, consistent with Tarski's material adequacy condition, the definition of truth in the case under consideration should be provided by the next T-sentence:\smallskip

\begin{equation} \label{CTSEN} %{Eq.71}
   S
   \mkern-2mu\text{ is true in }\mkern-2mu
   \mathcal{L}_{\mkern3mu\mathrm{INS}}
   \iff
   \mho_S
   \mkern-2mu\text{ is true in }\mkern-2mu
   \mathcal{L}_{\mkern3mu\mathrm{ZF}}
   \;\;\;\;  ,
\end{equation}
\smallskip

\noindent where $\mho_S\in\mathcal{L}_{\mkern3mu\mathrm{ZF}}$ is the set-theoretic counterpart of a sentence $S\in\mathcal{L}_{\mkern3mu\mathrm{INS}}$.\bigskip

\noindent It is straightforward to explain how classical logic arises from (\ref{CTSEN}). Certainly, let us introduce the set $X_S$ of states of the system that are associated with the sentence $S$. Then, $\mho_S$ can be taken as the formula $q\mkern-2mu\in\mkern-2muX_S$ evaluated to true for each given state $q$ of the system. Subsequently, it follows:\vspace{-8pt}

\begin{alignat}{3}
   S                          %{Eq.72}
   \text{ is true in }
   \mathcal{L}_{\mkern3mu\mathrm{INS}}
   &
   \iff
   q
   \mkern-2mu\in\mkern-2mu
   X_S
   \mkern-2mu\text{ is true in }\mkern-2mu
   \mathcal{L}_{\mkern3mu\mathrm{ZF}}
   &
   \;\;\;\;  ,
\\[8pt]
   \neg S                   %{Eq.73}
   \text{ is true in }
   \mathcal{L}_{\mkern3mu\mathrm{INS}}
   &
   \iff
   q
   \mkern-2mu\in\mkern-2mu
   X_{S}^{c}
   \mkern-2mu\text{ is true in }\mkern-2mu
   \mathcal{L}_{\mkern3mu\mathrm{ZF}}
   &
   \;\;\;\;  ,
\\[8pt]
   S_ 1\sqcap S_2      %{Eq.74}
   \text{ is true in }
   \mathcal{L}_{\mkern3mu\mathrm{INS}}
   &
   \iff
   q
   \mkern-2mu\in\mkern-2mu
   (X_{S_1} \mkern-2mu\cap\mkern-2mu X_{S_2})
   \mkern-2mu\text{ is true in }\mkern-2mu
   \mathcal{L}_{\mkern3mu\mathrm{ZF}}
   &
   \;\;\;\;  ,
\\[8pt]
   S_ 1\sqcup S_2      %{Eq.75}
   \text{ is true in }
   \mathcal{L}_{\mkern3mu\mathrm{INS}}
   &
   \iff
   q
   \mkern-2mu\in\mkern-2mu
   (X_{S_1} \mkern-2mu\cup\mkern-2mu X_{S_2})
   \mkern-2mu\text{ is true in }\mkern-2mu
   \mathcal{L}_{\mkern3mu\mathrm{ZF}}
   &
   \;\;\;\;  ,
\end{alignat}
\vspace{-8pt}

\noindent where $X_S^c = \{ x\in\mathbb{R}\mkern2mu|\mkern2mu x\notin X_S\}$ is the complement of $X_S$, $\cap$ and $\cup$ denote set-theoretic operations of intersection and union respectively. It can be said that sentences of the instrumentalist description of macroscopic phenomena have equivalent expressions in a Boolean algebra of sets. Therefore, as expressed by the language of classical mechanics $\mathcal{L}_{\mkern3mu\mathrm{CM}}$, instrumentalism obeys classical logic \cite{Halmos}.\bigskip

\noindent Yet, in the belief that quantum mechanics is a universal theory and so everything is ultimately describable in quantum-mechanical terms, one must insist that any sentence $S$ of the instrumentalist description of the physical world -- even at the macroscopic scale -- should be true if and only if the formula $\hat{P}_S |\Psi\rangle\mkern-3.5mu = \mkern-3.5mu |\Psi\rangle$, the copy of $S$ in the language of Hilbert space theory, is true. So, the following hierarchy of metalanguages must hold:\smallskip

\begin{equation} \label{RECD} %{Eq.76}
   S
   \mkern-2mu\text{ is true in }\mkern-2mu
   \mathcal{L}_{\mkern3mu\mathrm{INS}}
   \iff
   q
   \mkern-2mu\in\mkern-2mu
   X_S
   \mkern-2mu\text{ is true in }\mkern-2mu
   \mathcal{L}_{\mkern3mu\mathrm{ZF}}
   \iff
   \hat{P}_{S}
   |\Psi\rangle
   \mkern-3.5mu = \mkern-3.5mu
   |\Psi\rangle
   \mkern-2mu\text{ is true in }\mkern-2mu
   \mathcal{L}_{\mkern3mu\mathrm{HIL}}
   \;\;\;\;  .
\end{equation}
\smallskip

\noindent It is appeared from the above that the assumption of the universality of quantum mechanics requires the language of Hilbert space theory $\mathcal{L}_{\mkern3mu\mathrm{HIL}}$ to be a metalanguage with respect to the language of set theory $\mathcal{L}_{\mkern3mu\mathrm{ZF}}$:\smallskip

\begin{equation}  %{Eq.77}
   \phi
   \mkern-2mu\text{ is true in }\mkern-2mu
   \mathcal{L}_{\mkern3mu\mathrm{ZF}}
   \iff
   \psi_{\phi}
   \mkern-2mu\text{ is true in }\mkern-2mu
   \mathcal{L}_{\mkern3mu\mathrm{HIL}}
   \;\;\;\;  ,
\end{equation}
\smallskip

\noindent where $\phi$ stands for any WFF of the object language $\mathcal{L}_{\mkern3mu\mathrm{ZF}}$ that may be interpreted as representing a declarative sentence and $\psi_{\phi}$ is the copy of $\phi$ in the metalanguage $\mathcal{L}_{\mkern3mu\mathrm{HIL}}$.\bigskip

\noindent Alas, such a requirement cannot be met.\bigskip

\noindent To see this, first recall that a metalanguage must have primitive notions absent from an object language \cite{Murawski}. Next, consider the empty set $\{\mkern0.5mu\}$, a primitive notion in the language of set theory (for to assert that $\{\mkern0.5mu\}$ exists, one needs an axiom). The empty set cannot be a subspace of any Hilbert space $\mathcal{H}$, inasmuch as $\{\mkern0.5mu\}$ has no elements and a subspace of $\mathcal{H}$ contains at least the additive identity $\{0\}$ and so no less than one element.\bigskip

\noindent Consequently, the language of set theory has as a minimum one notion that cannot be expressed by any set of sentences in the language of Hilbert space theory. This entails that $\mathcal{L}_{\mkern3mu\mathrm{HIL}}$ cannot be a metalanguage with respect to $\mathcal{L}_{\mkern3mu\mathrm{ZF}}$.\bigskip

\noindent Therefore, mathematically, the hierarchy of metalanguages in (\ref{RECD}) is wrong.\bigskip

\noindent It implies, among other things, that classical mechanics cannot be regarded as the limiting case of quantum mechanics in the exact sense of the word. Certainly, because truth in set theory is not definable in the theory of Hilbert spaces (as an example, the true formula $\forall x\mkern-0.5mu{:}\mkern6mu x\mkern-2mu\notin\mkern-2mu\mkern-2mu\{\mkern0.5mu\}$ is not definable using vectors and subspaces of a Hilbert space $\mathcal{H}$), true formulas of classical mechanics cannot be described or determined using quantum mechanics. Because of that, the ``classical'' limit $\hbar\to 0$ of quantum mechanics cannot be identified with classical mechanics (in accordance with \cite{Klein}, the limit of quantum mechanics, as $\hbar$ approaches 0, is a classical statistical theory referred to as probabilistic Hamilton-Jacobi theory, not Newtonian mechanics).\bigskip

\noindent As it turns out, the simple mathematical fact that the empty set is not equal to a subspace of any vector space (the inequality $\{\mkern0.5mu\}\mkern-4mu\neq\mkern-4mu\{0\}$ can be used as a sign of this fact) calls into doubt Bohr’s correspondence principle and, in consequence of that, the reductionist approach to the description and understanding of natural phenomena.\bigskip

\section{Sequence of theories as the hierarchy of metalanguages}  %{<----------------------------------------------------------------------------------------Section VII}

\noindent It may be argued that the hierarchy of mathematical theories of natural sciences, wherein each theory could be derivable from more fundamental ones located on the top of it \cite{Taylor, Tegmark}, corresponds exactly to the hierarchy (i.e., the pre-ordered set) of their languages\smallskip 

\begin{equation} \label{HIER} %{Eq.78}
   \mathcal{L}_{\mkern3mu\mathrm{0}}
   \subseteq
   \mathcal{L}_{\mkern3mu\mathrm{1}}
   \subseteq
   \cdots
   \subseteq
   \mathcal{L}_{\mkern3mu\mathrm{\alpha}}
   \subseteq
   \mathcal{L}_{\mkern3mu\mathrm{\alpha +1}}
   \subseteq
   \cdots
   \;\;\;\;    
\end{equation}
\smallskip

\noindent wherein for every sentence at level $\alpha$ there is a sentence at level $\alpha + 1$ which asserts that the first sentence is true.\bigskip

\noindent The hierarchy (\ref{HIER}) may be finite or infinite. In the former case, it stops at the ultimate metalanguage or metatheory capable of defining truth for any other theory. As a result, this ultimate metatheory can be believed to be the singular, all-encompassing, coherent theoretical framework of nature, i.e., the so-called TOE, “theory of everything”.\bigskip
 
\noindent It is worth noting that unless the TOE might be allowed to be self-contradictory (inconsistent), it could not have the truth predicate. In more detail, let’s consider the following biconditional:\smallskip

\begin{equation}  %{Eq.79}
   \text{AtomSent}\mkern1.5mu\phi_{\alpha}
   \mkern-2mu\text{ is true in }\mkern-2mu
   \mathcal{L}_{\mkern3mu\mathrm{\alpha}}
   \iff
   \Phi_{\phi_{\alpha}}
   \mkern-2mu\text{ is true in }\mkern-2mu
   \mathcal{L}_{\mkern3mu\mathrm{TOE}}
   \;\;\;\;  ,
\end{equation}
\smallskip

\noindent where $\text{AtomSent}\mkern1.5mu\phi_{\alpha}$ expresses that $\phi_{\alpha}$, a WFF of the formal language $\mathcal{L}_{\alpha}$, can be interpreted as representing an atomic declarative sentence and $\Phi_{\phi_{\alpha}}$ is the unique copy of $\phi_{\alpha}$ in the ultimate metalanguage $\mathcal{L}_{\mkern3mu\mathrm{TOE}}$. At the topmost level of the hierarchy, this biconditional would become\smallskip

\begin{equation}  %{Eq.80}
   \text{AtomSent}\mkern1.5mu\phi_{\mkern2mu\mathrm{TOE}}
   \mkern-2mu\text{ is true in }\mkern-2mu
   \mathcal{L}_{\mkern3mu\mathrm{TOE}}
   \iff
   \Phi_{\phi_{\mkern2mu\mathrm{TOE}}}
   \mkern-2mu\text{ is true in }\mkern-2mu
   \mathcal{L}_{\mkern3mu\mathrm{TOE}}
   \;\;\;\;  .
\end{equation}
\smallskip

\noindent Since $\Phi_{\phi_{\mkern2mu\mathrm{TOE}}}$ is just $\text{AtomSent}\mkern1.5mu\phi_{\mkern2mu\mathrm{TOE}}$ itself, the above indicate that $\text{AtomSent}\mkern1.5mu\phi_{\mkern2mu\mathrm{TOE}}$ would be self-referential in $\mathcal{L}_{\mkern3mu\mathrm{TOE}}$. Let $\text{AtomSent}\mkern1.5mu\phi_{\mkern2mu\mathrm{TOE}}$ be false, then this sentence would be true if and only if it was not. The last can be presented as the liar paradox:\smallskip

\begin{equation}  %{Eq.81}
   \text{AtomSent}\mkern1.5mu\phi_{\mkern2mu\mathrm{TOE}}
   \mkern-2mu\text{ is true in }\mkern-2mu
   \mathcal{L}_{\mkern3mu\mathrm{TOE}}
   \iff
   \text{AtomSent}\mkern1.5mu\phi_{\mkern2mu\mathrm{TOE}}
   \mkern-2mu\text{ is not true in }\mkern-2mu
   \mathcal{L}_{\mkern3mu\mathrm{TOE}}
   \;\;\;\;  .
\end{equation}
\smallskip

\noindent To prevent the liar paradox from happening in the TOE – that is, to permit consistency of the TOE – the predicate ``is true (false)'' must not apply to atomic declarative sentences of $\mathcal{L}_{\mkern3mu\mathrm{TOE}}$.\bigskip

\noindent In this way, if existed, the TOE might be inconsistent or consistent. In the former case, it would be a decidable theory (for which there is an effective procedure that decides whether a given formula is a member of the theory or not). But in the latter case, the TOE would be a theory in which atomic declarative sentences could not be true or false. In consequence, the soundness of the consistent TOE would be an undefinable notion, which informally means that one would be unable to tell whether the inference rules of the consistent TOE were correct.\bigskip

\noindent As an alternative, in case of the infinite hierarchy (\ref{HIER}), our knowledge of the physical world would be forever incomplete. More importantly though, the infinite hierarchy would entail the existence of true, meaningful sentences about natural phenomena that would be above every theory and so impossible to define within any of them.\bigskip

\noindent The problem, however, is that the idea of arranging the universe of discourse (mathematical theories of natural sciences, their formal languages) in levels – whether finite or not – is incompatible with the inequality $\{\mkern0.5mu\}\mkern-4mu\neq\mkern-4mu\{0\}$. Furthermore, the fact that the language of Hilbert space theory $\mathcal{L}_{\mkern3mu\mathrm{HIL}}$ possesses the primitive notion of \emph{noncommutativity} absent from the language of set theory $\mathcal{L}_{\mkern3mu\mathrm{ZF}}$ exacerbates the matter.\bigskip

\noindent This suggests that there are WFFs in $\mathcal{L}_{\mkern3mu\mathrm{HIL}}$ which cannot be expressed in $\mathcal{L}_{\mkern3mu\mathrm{ZF}}$ as well as WFFs in $\mathcal{L}_{\mkern3mu\mathrm{ZF}}$ which cannot be expressed in $\mathcal{L}_{\mkern3mu\mathrm{HIL}}$. That is, the languages $\mathcal{L}_{\mkern3mu\mathrm{ZF}}$ and $\mathcal{L}_{\mkern3mu\mathrm{HIL}}$ are distinct, but neither is more expressive than the other. Consequently, $\mathcal{L}_{\mkern3mu\mathrm{ZF}}$ and $\mathcal{L}_{\mkern3mu\mathrm{HIL}}$ cannot be in a hierarchical relation. Providing the power of a theory is determined by the expressive power of its formal language, this can be understood that quantum mechanics cannot be more powerful than its classical counterpart.\bigskip

\noindent Such a conclusion flatly contradicts the usual assumed hierarchy of theories whereby quantum theory is more fundamental – and thus more powerful – than classical mechanics.\bigskip

\section{Possible solutions to the problem of the hierarchy of theories}  %{<------------------------------------------------------------------------------------Section VIII}

\noindent To preserve the concept of hierarchical organization of theories, one can assume that a formal language of quantum mechanics $\mathcal{L}_{\mkern3mu\mathrm{QM}}$ is not $\mathcal{L}_{\mkern3mu\mathrm{HIL}}$ but the one that is included in $\mathcal{L}_{\mkern3mu\mathrm{ZF}}$. In symbols,\smallskip

\begin{equation}  %{Eq.82}
   \mathcal{L}_{\mkern3mu\mathrm{QM}}
   \subseteq
   \mathcal{L}_{\mkern3mu\mathrm{ZF}}
   \;\;\;\;  .
\end{equation}
\smallskip

\noindent As $\mathcal{L}_{\mkern3mu\mathrm{CM}}$ is also included in $\mathcal{L}_{\mkern3mu\mathrm{ZF}}$, this assumption makes it is possible to put $\mathcal{L}_{\mkern3mu\mathrm{QM}}$ and $\mathcal{L}_{\mkern3mu\mathrm{CM}}$ in a hierarchical relation, that is, $\mathcal{L}_{\mkern3mu\mathrm{CM}}\subseteq\mathcal{L}_{\mkern3mu\mathrm{QM}}$ or $\mathcal{L}_{\mkern3mu\mathrm{QM}}\subseteq\mathcal{L}_{\mkern3mu\mathrm{CM}}$. The former conventionally implies that classical physics emerges from quantum theory, whereas the latter oppositely states that quantum theory is derivable from more fundamental classical theory.\bigskip

\noindent It must be acknowledged that the hierarchy $\mathcal{L}_{\mkern3mu\mathrm{QM}}\subseteq\mathcal{L}_{\mkern3mu\mathrm{CM}}$ has a long and respectable list of advocates, including among others Einstein, Schrödinger, and Bell. Most recent support for it has come from Leggett as well as from “spontaneous collapse” theorists such as Pearle, Ghirardi, Rimini, Weber, and others \cite{Landsman}. Furthermore, the said hierarchy agrees with the position of ’t Hooft claiming that quantum mechanics can be perfectly understood as a completely conventional, yet complex, classical theory \cite{Hooft}.\bigskip

\noindent Alternatively, given that the empty set $\{\mkern0.5mu\}$ cannot be in a Hilbert space, one may consider removing it from ZF as well. However, $\{\mkern0.5mu\}$ relates to the infinity axiom of ZF, $\mathsf{Inf}$, that reads in symbols:\smallskip

\begin{equation}  %{Eq.83}
   \mathsf{Inf}
   \mathrel{\mathop:}=
   \mkern3mu
   \exists\mkern2mu
   \mathbf{I}
   \Big(
      \{\mkern0.5mu\}
      \in
      \mathbf{I}
      \sqcap
      \forall x
      \left(
         x
         \in
         \mathbf{I}
         \to
         x
         \cup
         \{x\}
         \in
         \mathbf{I}
      \right)
      \mkern-2mu
   \Big)
   \;\;\;\;  ,
\end{equation}
\smallskip

\noindent or in words: “There is a set $\mathbf{I}$ (postulated to be infinite) such that the empty set $\{\mkern0.5mu\}$ is in $\mathbf{I}$, and such that whenever any $x$ is a member of $\mathbf{I}$, the set formed by taking the union of $x$ with its singleton $\{x\}$ is also a member of $\mathbf{I}$” \cite{Karel}. More colloquially, the axiom $\mathsf{Inf}$ introduces the notion of infinity to ZF that allows an endless procedure to be regarded as a whole, a single element in further constructions \cite{Baratella}. And forasmuch as $\{\mkern0.5mu\}$ is the member of the set guaranteed to exist in the infinity axiom $\mathsf{Inf}$, to get rid of $\{\mkern0.5mu\}$ one must remove $\mathsf{Inf}$.\bigskip

\noindent Thus, an alternative set theory, AST, whose power is comparable to that of Hilbert space theory may symbolically be presented as\smallskip

\begin{equation} \label{AST} %{Eq.84}
   \mathrm{AST}
   \mathrel{\mathop:}=
   \mkern3mu
   \mathrm{ZF}
   -
   \mathsf{Inf}
   +
   \neg\mathsf{Inf}
   -
   \mathsf{Spec}
   +
   \overline{\mathsf{Spec}}
   +
   \mathscr{X}
   \;\;\;\;  ,
\end{equation}
\smallskip

\noindent where $\mathsf{Spec}$ is the axiom schema of specification of ZF, $\overline{\mathsf{Spec}}$ is its restricted version (that denies the existence of intersection of disjoint sets, namely, according to $\overline{\mathsf{Spec}}$, if any two sets do not have any common element, then they have no intersection), and $\mathscr{X}$ represents axioms of AST nonexistent in ZF.\bigskip

\noindent Consequently, one can write down the following hierarchy:\smallskip

\begin{equation}   %{Eq.85}
   \mathcal{L}_{\mkern3mu\Gamma_{n}}
   \subseteq
   \mathcal{L}_{\mkern3mu\mathbb{C}^{n}}
   \subseteq
   \mathcal{L}_{\mkern3mu\mathrm{AST}}
   \;\;\;\;    ,
\end{equation}
\smallskip

\noindent where $\mathcal{L}_{\mkern3mu\Gamma_{n}}$ is a formal language over a classical phase space $\Gamma_{n}$, which has only finitely many elements, and $\mathcal{L}_{\mkern3mu\mathbb{C}^{n}}$ is a formal language over a finite-dimensional Hilbert space $\mathbb{C}^{n}$ with the inner (scalar) product $\langle z,z^{\prime}\rangle = \sum_{j=1}^{n} z_j^{\ast} z^{\prime}_j$.\bigskip

\noindent As members of the language of a theory of finite sets $\mathcal{L}_{\mkern3mu\mathrm{AST}}$, WFFs of $\mathcal{L}_{\mkern3mu\Gamma_{n}}$ and $\mathcal{L}_{\mkern3mu\mathbb{C}^{n}}$ have no singularities. This should enable classical and quantum theories to give physically meaningful results without modifying observables or assuming new physics.\bigskip

\noindent Moreover, inside AST axiomatized by (\ref{AST}), any geometric system has only a finite number of points and for that reason is guaranteed from infinities arising in calculated quantities \cite{Suppes}. Thus, as long as a geometric theory of gravitation can be expressed in a formal language over a finite geometry, gravity may be treated as a quantum field. That is, quantum gravity may be characterized by a choice of finitely many parameters which could in principle be set by experiment, implying that a theory of quantum gravity can make predictions.\bigskip

\noindent Still and all, one may object to the negation of the axiom of infinity $\mathsf{Inf}$ proposed in (\ref{AST}) saying that mathematics without infinite elements, i.e., $+\infty$ and $-\infty$, is too restrictive and gives up its important part. As a reply to this objection, it can be pointed out that even in mathematics an infinite element cannot be accepted as an outcome because, as all computational resources are finite, it would be that these resources have been used up before the outcome $\pm\infty$ has been reached \cite{Bendegem}.\bigskip

\noindent Another thing, nowadays there is no evidence that Zermelo-Frankel set theory needs any change to settle open problems of prima facie mathematical interest \cite{Feferman99}. However, as it follows from the above discussion, such a change might be a solution to the problem of the hierarchy of physics theories, particularly, the incompatibility between general relativity and quantum mechanics.\bigskip

\section*{Acknowledgement}  %{<-------------------------------------------------------------------------------------------------}

\noindent The author wishes to thank the anonymous referee for the productive comments and interesting remarks that helped offset shortcomings of a draft of this paper.\bigskip

\bibliographystyle{References}

\end{document}